# Interaction of Polar and Nonpolar Organic Pollutants with Soil Organic Matter: Sorption Experiments and Molecular Dynamics Simulation

Ashour A. Ahmed*[a,b], Sören Thiele-Bruhn[c], Saadullah G. Aziz[d], Rifaat H. Hilal[b,d], Shaaban A. Elroby[d,e], Abdulrahman O. Al-Youbi[d], Peter Leinweber[f], Oliver Kühn[a]

[a] University of Rostock, Institute of Physics, D-18051 Rostock, Germany
[b] University of Cairo, Faculty of Science, Department of Chemistry, 12613 Giza, Egypt
[c] University of Trier, Soil Science, D-54286 Trier, Germany
[d] King Abdulaziz University, Faculty of Science, Department of Chemistry, Jeddah 21589, Saudi Arabia
[e] University of Beni-Suef, Faculty of Science, Department of Chemistry, Beni-Suef, Egypt
[f] University of Rostock, Soil Science, D-18051 Rostock, Germany

ashour.ahmed@uni-rostock.de       (The corresponding Author)

thiele@uni-trier.de
saziz@kau.edu.sa
rhilal@kau.edu.sa
skamel@kau.edu.sa
aalyoubi@kau.edu.sa
peter.leinweber@uni-rostock.de
oliver.kuehn@uni-rostock.de

# Abstract

The fate of organic pollutants in the environment is influenced by several factors including the type and strength of their interactions with soil components especially SOM. However, a molecular level answer to the question "**How organic pollutants interact with SOM?**" is lacking. In order to explore mechanisms of this interaction, we have developed a new SOM model followed by carrying out molecular dynamics (MD) simulations in parallel with sorption experiments. The new SOM model comprises free SOM functional groups (carboxylic acid and naphthalene) as well as SOM cavities (with two different sizes), representing the soil voids, containing the same SOM functional groups. To examine the effect of the hydrophobicity on the interaction, the organic pollutants hexachlorobenzene (HCB, non-polar) and sulfanilamide (SAA, polar) were considered. The experimental and the theoretical outcomes explored four major points regarding sorption of SAA and HCB on soil. 1- The interaction depends on the SOM chemical composition more than the SOM content. 2- The interaction causes a site-specific adsorption on the soil surfaces. 3- Sorption hysteresis occurs, which can be explained by inclusion of these pollutants inside soil voids. 4- It was observed that the hydrophobic HCB is adsorbed on soil stronger than the hydrophilic SAA. Moreover, the theoretical results showed that HCB forms stable complexes with all SOM models in the aqueous solution while most of SAA complexes are accompanied by dissociation. The SOM-cavity modeling showed a significant effect on binding of organic pollutants to SOM. Both HCB and SAA bind to the SOM models in the order of models with a small cavity > models with a large cavity > models without a cavity. Although HCB binds to all SOM models stronger than SAA, the latter is more affected by presence of the cavity. Finally, HCB and SAA bind to the hydrophobic functional group (naphthalene) stronger than to the hydrophilic one (carboxylic acid) for all SOM models containing cavity. For models without a cavity, SAA binds to carboxylic acid stronger than to naphthalene.

# 1 Introduction

Soil is a complex heterogeneous mixture of mineral and organic matter in addition to water and air. Despite the low percentage of the soil organic matter (SOM) in most soils, it has substantial influence on soil properties and agricultural productivity. As the current state of knowledge, SOM is regarded as highly polydisperse, supra-molecular assemblies, characterized by multiple constituents and functional groups, voids and variable rigidity (Schulten and Schnitzer, 1998; Schaumann, 2006; Senesi et al., 2009). SOM consists of carbohydrates, peptides, alkyl aromatics, N-heterocyclic compounds, phenols, sterols, lignins, lipids, hydrocarbon, bound and free fatty acids, nitriles, suberin to name just the most important compounds (Schulten and Leinweber, 1999). SOM largely governs soil functions, one being the retardation of pollutants in soil.

Persistent organic pollutants (POPs) are among the most environmentally hazardous compounds. Due to their hydrophobicity, they are resistant to environmental degradation through chemical, biological, and photolytic processes and are tending to accumulate in soil and ground water (Ritter et al., 1995). They are ubiquitously distributed in the environment having long life times of up to years and decades in the soil (Jones and de Voogt, 1999). Additionally, polar organic pollutants such as pharmaceuticals and personal care products, have been recognized as emerging soil pollutants in the past decades. They reach soil through contaminated wastewater and biosolids as well as excreta from medicated livestock used as fertilizer (Boxall et al., 2004; Jjemba, 2006). In general, they are mostly polar and ionizable compounds, bearing various functional groups (Thiele-Bruhn, 2003). The fate of the different types of pollutants in the environment is influenced by several factors. The most important one is the type and strength of interactions of the pollutant to soil components especially SOM. Consequently, the physical and chemical properties of the soil as well as the pollutant play a pivotal role to explore nature of pollutant-soil interaction.

The extent and concentration dependence of pollutant-SOM interaction can be investigated via adsorption experiments of the pollutant on soil. Adsorption of hydrophobic pollutants on soils is described by two stage kinetics (Weber et al., 1991; Chen et al., 2004). The vacant sites are filled up in a rapid initial stage while diffusion of the hydrophobic pollutants into SOM is a slow rate-limited process (Chiou et al., 1983). In contrast, the mechanism of sulfonamide adsorption may include hydrogen bonding, van der Waals forces, cation exchange, cation bridging, and surface complexes (MacKay and Canterbury, 2005; Lertpaitoonpan et al., 2009) in addition to hydrophobic interaction. Their sorption is strongly affected by pH and ionic strength (Ter Laak et al., 2006; Gao and Pedersen, 2005; Kurwadkar et al., 2007). Since adsorption experiments yield information that can only be correlated statistically to soil properties, combination of molecular modeling and computational chemistry is a complementary promising approach to develop an atomistic understanding of the binding of pollutants to soil (Gerzabek et al., 2001; Schaumann and Thiele-Bruhn, 2011).

Due to the heterogeneity and complexity of SOM composition, still its modeling is a complicated problem and the proposed models have not been introduced without triggering criticism. There are different hypotheses concerning the SOM principal structural organization (Schaumann and Thiele-Bruhn 2011), i.e. macromolecular vs. supramolecular (Piccolo 2002; Sutton and Sposito 2005; Schaumann 2006). The first hypothesis for a polymeric SOM model has been developed by Schulten and coworkers on the basis of bio- and geochemical, NMR-spectroscopic and mass spectrometric analyses (Schulten and Schnitzer 1995; Schulten and Leinweber 2000; Schulten 2002). It was established by creating a network including aromatic rings through long chain alkyl structures (Schulten et al., 1991). Moreover, some carbohydrate and protein units can be inserted into this model in its internal voids and on its surfaces (Schulten and Schnitzer, 1993, 1997). This modeling approach could be criticized because of the huge number of possibilities for combining all SOM compounds and functional groups together into a single macromolecule. These models have been used for molecular mechanics calculation of the interaction of POPs as well as polar organic pollutants with SOM. Results showed that hydrophobic chemicals are especially retarded through hydrophobic interactions, e.g. $\pi-\pi$ bonds and alignment of aromatic rings, while polar sorbates interacted most through H-bonds and van der Waals forces (Schulten and Violante, 2002). Moreover, trapping in microvoids of the flexible SOM structure was identified as a relevant mechanism for the immobilization of both polar and hydrophobic organic sorbates that is often, though not necessarily combined with additional binding forces (Schulten and Leinweber, 2000). For example, computational modeling confirmed that entrapment in voids of SOM is crucial for the sorption of pharmaceutical antibiotic sulfonamides in soil (Schwarz et al., 2012; Thiele-Bruhn et al., 2004). However, the role of voids does not always match with experimental findings (e.g., Wang and Xing, 2005) and needs further investigation.

Another approach is describing SOM as a set of SOM functional groups (Aquino et al., 2007, 2009). Here, a deficiency could arise from modeling of SOM by few numbers of functional groups only. Therefore, to overcome these problems, recently Ahmed et al. (2014a, 2014b) have developed an advanced approach for SOM modeling based on an experimental SOM characterization by different analytical techniques (Ahmed et al., 2012). This model includes a large test set of separate representative systems covering the most relevant functional groups as well as analytically determined compound classes. The validity of this model has been checked by experimental adsorption of hexachlorobenzene on well-characterized soil samples (Ahmed et al., 2014a). In these studies it had been stressed that further improvement and testing of this model will be needed, especially in case of explicit solvation by water molecules.

The objective of the present study was to model the sorption of the hydrophobic hexachlorobenzene (HCB) and of the polar pharmaceutical sulfonamide sulfanilamide (SAA) by molecular dynamics to explore general aspects of the interaction of both the hydrophobic and hydrophilic organic pollutants to SOM. We did this by using the above selected, representative sorptive systems of a previously developed SOM model test set (Ahmed et al., 2014a,b), now for the first time incorporating the effect of void-restricted interactions and molecular dynamics simulations in the presence of water.

## 2 Materials and methods

### 2.1 Adsorbates

Two organic compounds from different organic pollutant categories were selected in the current study, i.e. HCB with $\log K_{OW}$ of 4.89 and SAA with $\log K_{OW}$ of -0.62. The three-dimensional geometries for both HCB and SAA are given in Fig. 1. Results on soil adsorption of HCB have been recently published (Ahmed et al., 2014a), hence, information on sorption experiments is focused on SAA adsorption. SAA was obtained from Sigma (Taufkirchen, Germany) with purity of ≥99.0%.

**Figure 1**

### 2.2 Soil site and sampling

Three topsoil samples were obtained from an unfertilized and two differently fertilized plots of the long-term Eternal Rye Cultivation experiment at Halle (Saale), Germany (Kühn, 1901). Representative samples of the haplic Phaeozem Ap horizon (0–20 cm depth) were taken by a corer from the unfertilized treatment (U), a treatment that received farmyard manure in the time period 1878 to 1963 and remained unfertilized since then (StmII), and a treatment that received farmyard manure from 1878 until sampling date in 2000 (StmI). These samples were air dried and sieved (<2 mm), and analyzed for basic properties (Table 1). All three samples had a similar mineral composition with illite, smectite and mixed layer minerals predominating in the clay fraction (< 2 μm) (Leinweber and Reuter, 1989).

### 2.3 Adsorption isotherms of SAA on soil samples

Adsorption isotherms were determined in batch trials based on previous studies (e.g., Thiele-Bruhn et al., 2004). All samples were prepared in triplicate. For each replicate, 5.0 g of air-dried soil sample was weighed into 75-mL glass centrifuge tubes and spiked with SAA dissolved in methanol. The concentrations added (0, 0.1, 1, 10, and 40 mg/kg) were selected to represent the predicted environmental concentrations of sulfonamides. After the solvent was allowed to evaporate for 1 h, 0.01 $M$ CaCl$_2$ was added in a soil to solution ratio of 1:2.5 (w/v). Samples were shaken on an end-over-end rotary shaker at 15 rpm for 16 h at 22°C in the dark. Subsequently, they were centrifuged for 30 min at 1700 ×$g$.

Desorption was investigated following the adsorption step by replacing the supernatant after centrifugation and decanting with the equivalent volume of methanol. Subsequently, samples were shaken and further treated as described for the adsorption step. The SAA recovered in methanol is defined as the "total desorbable fraction" (Thiele-Bruhn and Aust, 2004).

Supernatants of CaCl$_2$ and methanol from adsorption and desorption experiments, respectively, were directly analyzed. Gradient HPLC analysis was performed on a Hewlett-Packard (Palo Alto, CA) 1050 HPLC system equipped with a wavelength programmable UV detector (HP 1050) and a fluorescence detector (HP 1046A). A 250- × 4.6-mm, reversed-phase Nucleosil 100-5-C18 column was used (Macherey & Nagel, Düren, Germany). The mobile phase consisted of 0.01 $M$ H$_3$PO$_4$ and methanol with a flow rate of 1.0 mL min$^{-1}$. Using injection volumes of 10 μL, SAA was determined with UV detection at 265 nm and fluorescence detection at 276/340 nm.

The nonlinear Freundlich isotherm (Eq. 1) was fitted to the data of the total sorbate concentration associated with the sorbent (C$_s$, μmol kg$^{-1}$) and the total chemical concentration remaining in the equilibrium solution (C$_w$, μmol L$^{-1}$) using the CFIT software for nonlinear regression (Helfrich, 1996):

$$C_s = K_F \cdot C_w^n \tag{1}$$

where K$_F$ is the Freundlich unit capacity (μmol$^{1-n}$ L$^n$ kg$^{-1}$) and n is the Freundlich exponent which is a measure of nonlinearity.

### 2.4 SOM modeling

For the description of binding of organic pollutants to SOM, a new SOM model has been developed recently (Ahmed et al., 2014a) based on detailed molecular analyses by pyrolysis field ionization mass spectrometry (Py-FIMS) and X-ray absorption near edge structure spectroscopy (XANES). The key aspect of this SOM model is the consideration of separate representative systems, covering carbohydrates, phenols and lignin monomers, lipids, alkyl aromatics, N-containing compounds including peptides, and other compounds containing different functional groups (Ahmed et al., 2012). The validity of this model has been verified through correlation of the calculated binding energies of its components to HCB with the outcome of adsorption measurements of HCB on different soil samples. So far the model had been restricted to a continuum solvation approach. However, the semi-quantitative agreement between quantum chemical and molecular mechanics force field interaction energies (Ahmed et al., 2014b), suggests an extension of the model to include explicit water molecules. Initial investigations revealed that HCB-SOM binding competes with solvation of the separate SOM and HCB moieties. This points to another deficiency of our model, i.e. the missing effect of the supramolecular arrangement of the SOM particle, providing a heterogeneous surface as well as cavities for binding. Both issues will be addressed here by combining a simple model for the supramolecular structure with explicit solvating water molecules. Specifically, SOM will be modeled as a cyclic long aliphatic chain, forming a cavity that contains one functional group attached to the cavity through a short aliphatic chain (see Fig. 2c-f). To avoid complications due to the structural flexibility of the cavity, it is constructed from a long chain-conjugated alkene (46 carbon atoms). Out of the functional groups studied in Ahmed et al. (2014a), two have been selected for the present investigation, namely naphthalene (Fig. 2a, c, and e) and carboxylic acid (Fig. 2b, d, and f), which represent extreme cases concerning the interaction (binding energy) with HCB (Ahmed et al., 2014a). In addition, these two cases represent nonpolar and polar functional groups. This cavity model may account for trapping and binding of organic pollutants in the spirit of the void approach by Schulten and coworkers (Schulten et al., 1991). Moreover, it can partially shield the SOM functional group and the pollutant from the surrounding water. Whether this shielding is effective or independent of the size of the cavity will be investigated by considering a second cavity made up by 90 carbon atoms (Fig. 2e-f).

To sum up this yields the following set of models: The bare SOM models studied previously will be called **Ia** for naphthalene (Fig. 2a) and **IIa** for carboxylic acid (Fig. 2b). The case of naphthalene in the small and large cavity will be labelled **Ib** (Fig. 2c) and **Ic** (Fig. 2e), respectively. Correspondingly, carboxylic acid within the small and the large cavity will be called **IIb** (Fig. 2d) and **IIc** (Fig. 2f), respectively.

For these setups averaged binding energies between the relevant pollutants (HCB and SAA) and the SOM models will be calculated using molecular dynamics simulations. Thereby attention is focused on effects due to (i) the presence of the SOM cavity, (ii) the different SOM functional groups included in the cavity, and (iii) the cavity size.

Figure 2

## 2.5 Molecular dynamics simulations

Initial structures of the pollutant-SOM complexes have been obtained by inserting the two pollutants, HCB and SAA, into the SOM models close to the naphthalene or carboxylic acid functional group. Each complex has been solvated in a cubic water box of 1 g.cm$^{-3}$ density, see Fig. 3, with edge length of 60 Å in case of **Ia** and **IIa**, 70 Å in case of **Ib** and **IIb**, and 90 Å in case of **Ic** and **IIc**. For the water, the simple point charge (SPC) model has been used. For all cases, energy minimization was followed by 100 ps equilibration runs (cf. Fig. 3). The subsequent 1 ns production runs have been performed using a canonical (NVT) ensemble, where the temperature (300 K) and the box volume are fixed. The time step was chosen as 2 fs. These calculations have been performed using the GROMACS program package (Hess et al., 2008) combined with the general CHARMM force field (Vanommeslaeghe et al., 2010).

For every system, the binding energy between the relevant pollutant (HCB or SAA) and the SOM functional group (naphthalene or carboxylic acid) has been calculated along the production trajectory. In addition, the binding energy is calculated between the pollutant and the SOM cavity in case of the systems including a cavity (**Ib, IIb, Ic,** and **IIc**). The calculated binding energy comprises contributions due to van der Waals and electrostatic interactions. The van der Waals interaction is described by the Lennard-Jones potential. The electrostatic interaction is represented by three contributions including Coulomb interaction, Poisson-Boltzmann reaction field interaction, and distance-independent reaction field interaction. The total binding energy ($E_B$) between two interacting sub-systems is expressed as follows:

$$E_B = \sum_{pairs\ i,j} \left[ \frac{C_{12}(i,j)}{r_{ij}^{12}} - \frac{C_6(i,j)}{r_{ij}^6} \right] + \sum_{pairs\ i,j} \frac{q_i q_j}{4\pi\epsilon_0 \epsilon_1} \left[ \frac{1}{r_{ij}} - \frac{C_{rf}\ r_{ij}^2}{2R_{rf}^3} - \frac{1 - \frac{1}{2}C_{rf}}{R_{rf}} \right]. \quad (2)$$

The first term represents the Lennard-Jones interaction, while the second one represents the total electrostatic interaction. The sums run over all non-bonded atom pairs i and j of the interacting sub-systems, where the parameters $C_6$ and $C_{12}$ depend on the type of atoms, $r_{ij}$ is the distance between the atoms, and the $q_i$ and $q_j$ are their partial charges. Further, $\epsilon_0$ and $\epsilon_1$ are the dielectric permittivity of the vacuum and the relative permittivity of the medium, $C_{rf}$ is the medium reaction-field constant, and $R_{rf}$ is a cutoff distance (Vanommeslaeghe et al., 2010).

Figure 3

## 3  Results and discussion

### 3.1  Sorption experiments

The long-term different treatments resulted in different organic carbon (OC) contents of the soils increasing in the order "U" < "StmII" < "StmI" (Table 1), which agrees with previous reports on this experiment (e.g., Schmidt et al., 2000). Sorption of the polar sulfonamide SAA to these three soil samples was best-fitted using the Freundlich isotherm (r² ≥ 0.85; Table 1). Due to the pH of the soil samples, sorption was fully restricted to the neutral SAA molecule (neutral species fraction ≥99.99 %). The Freundlich sorption coefficient ($K_F$) of SAA increased in the order "U" < "StmI"< "StmII". This indicates that adsorption of SAA on soil does not fully depend on the soil organic matter content. It is well known that soil sorption of polar sorbates such as sulfonamides also occurs to mineral colloids, although soil organic matter is most important (Essington et al., 2010; Figueroa-Diva et al., 2010; Gao and Pedersen 2005; Sanders et al., 2008; Thiele-Bruhn et al., 2004). Even more, sorption of SAA depends on the composition and properties of SOM (Thiele-Bruhn et al., 2004) that is clearly different between the differently treated plots of the long-term fertilization experiment, while the mineral composition of the three soils tested was not significantly different. The same outcome was explored for other sets of soil and particle-size samples experimentally and theoretically (Ahmed et al., 2014a; Thiele-Bruhn et al., 2004). The clear non-linearity of the obtained isotherms emphasizes that site-specific sorption occurred. This indicates that sorption mechanism and strength depended on the available functional groups within the SOM. This corresponds with previous studies on soil sorption of sulfonamides (Białk-Bielińska et al., 2012; Sanders et al., 2008; Thiele-Bruhn et al., 2004). While non-linearity was similar for "U" and "StmII" (Freundlich exponent n = 0.8), it was substantially smaller for "StmI" (Freundlich exponent n = 0.7). Hence, the difference in SOM composition, more than the differences in SOM quantity governed the difference in SAA specific sorption to the three tested soil samples. The dependence of sulfonamide sorption to soil on specific molecular sites and functional groups of SOM has been previously shown (Tolls, 2001; Thiele-Bruhn et al., 2004).

**Figure 4**

Upon desorption, the sorption coefficients substantially increased (Table 1) so that the ratio of $K_{F,sorption}/K_{F,desorption}$ was ≤0.48, showing strong sorption hysteresis. Such hysteresis of sulfonamide sorption has been previously reported and was interpreted to show energetically favorable specific sorption and inclusion of sulfonamides in voids of soil sorbents (Kahle and Stamm, 2007; Förster et al., 2009; Schwarz et al., 2012). Computational chemistry calculations using molecular mechanics (MM+), however, yielded contradicting results, showing entrapment of SAA in voids of a DOM trimer model (Schwarz et al., 2012), while preferred surface bonding was calculated for SAA using a different SOM model (Thiele-Bruhn et al., 2004). Hence, considering the limitations of molecular mechanics and the MM+ force field, it appeared as a specific task of this study to investigate the role of voids in SOM on the sorption of sulfonamides.

Recently, the adsorption of HCB on similar unfertilized soil sample (Ahmed et al., 2012) was investigated and a Freundlich adsorption coefficient ($K_F$) and exponent (n) of 4.02 and 0.75, respectively, have been reported (Ahmed et al., 2014a). Coinciding with the higher log$K_{ow}$ of HCB of 4.89 (log$K_{ow}$ SAA: -0.62), the sorption coefficient for HCB was substantially larger, indicating the higher strength of HCB sorption compared to SAA on similar samples. This resulted in an even higher adsorption of HCB to the unfertilized soil sample than of SAA to the fertilized soil samples, even though the latter have substantially higher organic matter content than the unfertilized soil sample.

**Table 1**

### 3.2 Simulation results

Based on the 1 ns production molecular dynamics trajectory sampling (canonical NVT ensemble), the interaction of both HCB and SAA (Fig. 1) with the different SOM models in aqueous solution has been analyzed. Here, the focus was on the binding energy and geometry of each pollutant with respect to the sub-units of the different SOM models. An overall picture of the thermal fluctuation dynamics is provided in Figures 5 and 6, which shows the pollutant-SOM configurations for selected samples along the 1 ns trajectory. Notice that for some systems, the pollutant did not bind to the SOM model along the whole trajectory. We considered the complex as dissociated if the interaction between the sub-systems was zero.

The averaged values of the following quantities have been calculated on the basis of the production trajectory: (i) the binding energies between HCB or SAA and the SOM functional group as well as the SOM cavity for each SOM model. (Table 2 and Table 4), (ii) the distance between the two centers of masses of the pollutant and the SOM functional group (Table 3 and Table 5), (iii) the radial distribution function for water molecules around each pollutant surface as well as the number of water molecules in a sphere with a radius of 5 Å around each pollutant.

#### 3.2.1 HCB-SOM interactions

Figure 4a shows that HCB is mainly bound to naphthalene **Ia** along the trajectory. The separating distance between HCB and naphthalene ranges from 3.3 to 9.3 Å with an average of 4.6 Å. The interaction between HCB and naphthalene is rather strong, with -26.1 kJ/mol as total averaged binding energy. Most of this energy, that is -25.4 kJ/mol, is coming from the van der Waals interaction. The contribution of Coulomb interactions is negligible (-0.7 kJ/mol). It was determined that on average 4.5 water molecules surround HCB in a sphere with a radius of 5 Å, whereas naphthalene is surrounded by 5.7 water molecules.

In comparison, HCB is less bound to the carboxylic acid group **IIa** than to naphthalene (Fig. 5b). The distance between HCB and the carboxylic acid group ranges from 3.5 to 47.2 Å with an average of 12.4 Å. In fact the HCB-SOM complex can be considered as being dissociated in the last quarter of the sampling trajectory. Before that phase the HCB-carboxylic acid group interaction oscillates between weak and strong binding. The average of the total binding energy of HCB with carboxylic acid group is -14.9 kJ/mol, leaving aside the dissociated phase. The van der Waals energy contributes with -15.0 kJ/mol while the contribution of the Coulomb interaction is +0.1 kJ/mol. In this complex, HCB is surrounded by 5.7 water molecules while the carboxylic acid group is surrounded by 8.8 water molecules. This indicates stronger solvation, longer distance separation, and consequently weaker binding energy in this complex compared to the HCB-naphthalene complex.

Next, we consider the effect of a SOM cavity. Figure 5c shows that HCB is strongly bound to naphthalene and stays within the SOM cavity along the whole trajectory. Thereby both HCB and naphthalene are mainly found in the plane of the SOM cavity (see Fig. S1a in the supplementary material). Due to the cavity effect, the distance between HCB and naphthalene changes from 3.4 to 10.6 Å with an average of 4.6 Å. Along the whole trajectory, there is a strong interaction for HCB with naphthalene and the SOM cavity. The average binding energy for HCB with this whole SOM model is -64.9 kJ/mol. This binding energy is subdivided into two contributions; one of them is related to binding of HCB to the SOM functional group and the other is related to binding of HCB to the SOM cavity. The respective binding energies are -28.1 kJ/mol and -36.8 kJ/mol, respectively This indicates that the presence of the cavity will increase binding of HCB to the SOM functional group compared to the free SOM functional group case. In the presence of the SOM cavity, HCB is surrounded by 1.6 water molecules while naphthalene is surrounded by 4.2 water molecules. Thus, as compared with the solvation of HCB and naphthalene without the cavity (Fig. 5a), it becomes evident that the cavity can shield both HCB and naphthalene from water molecules. Consequently, this leads to stronger interaction of both molecules.

In contrast to the naphthalene case, HCB is highly mobile during its interaction with the carboxylic acid functional group and the SOM cavity (Fig. 5d). Despite of the fact that carboxylic acid stays in the plane of the SOM cavity along the trajectory, HCB leaves the SOM cavity and ends up out of plane of that cavity (see Fig. S1b in the supplementary material). Compared with the interaction of HCB with model **Ib**, a longer distance is calculated between HCB and carboxylic acid altering from 3.5 to 13.7 Å with an average of 7.8 Å. Weak interaction is observed between HCB and the carboxylic acid along the trajectory except in some regions having no interaction at all. However, HCB strongly interacts with the SOM cavity during the whole trajectory. The average binding energy for HCB with the whole model is -38.5 kJ/mol, while that for HCB with the carboxylic group is -3.3 kJ/mol and with the SOM cavity is -35.2 kJ/mol. This shows that binding of HCB to the carboxylic acid functional group is weaker in the presence of the cavity compared to the free carboxylic acid functional group. Furthermore, HCB is surrounded by 3.8 water molecules while carboxylic acid is surrounded by 11.4 water molecules. This documents that changing the functional moiety from naphthalene to the carboxylic acid group increases solvation of the complex' sub-units, which consequently decreases the interaction strength with HCB. Moreover, including the cavity will shield HCB from water, but increases solvation of the carboxylic acid.

Having studied the principal effect of a cavity we turn to the question how this is influenced by the size of the cavity. Regarding model **Ic**, HCB is bound to naphthalene and a segment of the large SOM cavity around the naphthalene functional group (Fig. 5e). During the trajectory, both HCB and the naphthalene functional group move simultaneously in one direction, out of plane of the SOM cavity (see Fig. S1c in the supplementary material). Hence, HCB is separated from naphthalene by 3.4 to 10.5 Å with an average of 5.8 Å. Along the whole trajectory, there is a relatively strong interaction for HCB with naphthalene and the SOM cavity. In the first 500 ps of the trajectory, interaction of HCB with naphthalene is stronger than that with the SOM cavity but this behavior is reversed in the last 500 ps. The average binding energy for HCB with this whole model (**Ic**) is -44.3 kJ/mol. The binding energy of HCB with naphthalene is -19.8 kJ/mol while that with the SOM cavity is -24.5 kJ/mol. This indicates that the larger SOM cavity the lower binding of HCB to both naphthalene and the cavity. In comparison with **Ia**, binding of HCB to naphthalene itself decreases, but the effect of the cavity overcomes this change yielding high binding energy. Further, for model **Ic** on average 3.7 water molecules surround HCB while 4.4 water molecules surround naphthalene. This means that as the cavity size increases the solvation increases, but still it is lower than in case of absence of the cavity (**Ia**). Consequently, this leads to intermediate binding energy between in case of including of a small cavity and absence of the cavity.

For model **IIc**, HCB interacts with a small part of the cavity around the carboxylic acid functional group (Fig. 5f). HCB and the carboxylic acid functional group move simultaneously in two different directions, up and down, out of plane of the SOM cavity (see Fig. 1d in the supplementary material). This suggests some repulsion between HCB and **IIc** which is in contrast to the attraction between HCB and **Ic**. They are separated by an average distance of 8.8 Å, which varies from 3.6 to 19.9 Å. The average binding energy for HCB with model **IIc** is -33.5 kJ/mol in which -3.1 kJ/mol is related to the carboxylic acid while -30.4 kJ/mol is due to the SOM cavity. In comparison with model **IIb**, one can observe that increasing of the cavity size will diminish binding of HCB to the SOM cavity without significant effect on the carboxylic acid. The binding of HCB to the carboxylic acid itself is decreased compared to the free the carboxylic acid (**IIa**) but in general binding of HCB to **IIc** is higher than that to **IIa**. Furthermore, it is observed that HCB is surrounded by 4.1 water molecules while the carboxylic acid is surrounded by 10.7 water molecules. This means that the order of HCB solvation by water molecules is in case of small cavity (**IIb**) < large cavity (**IIc**) < without cavity (**IIa**). In contrast, it was found that solvation of the carboxylic acid has the order of without cavity (**IIa**) < large cavity (**IIc**) **<** small cavity (**IIb**). Similar to the case **Ic**, solvation will lead to intermediate binding energy between the cases of small cavity and absence of the cavity.

**Figure 5**

### 3.2.2 SAA-SOM- interactions

Having studied the effects of the cavity, its size, and its interior functional groups, now we investigate how these findings are influenced by the type of pollutant. For the interaction of SAA with naphthalene (**Ia**), it is found that SAA is only weakly bound to naphthalene (Fig. 6a). Along the trajectory, the distance between SAA and naphthalene varies from 3.6 to 40.4 Å with an average of 20.6 Å. Mainly the interaction of SAA with naphthalene is limited to the first half period of the trajectory before dissociation of the SAA-naphthalene complex takes place. Within the first part, SAA is separated from naphthalene by an average distance of 8.1 Å, which varies between 3.6 and 18.4 Å. The corresponding average binding energy of SAA with naphthalene is -8.5 kJ/mol. SAA is surrounded by 9.6 water molecules while naphthalene is surrounded by 8.2 water molecules. Regarding the interaction of SAA with carboxylic acid (**IIa**), it is observed that SAA is a little bit more bound to the carboxylic acid (see Fig. 6b) than that in case of **Ia** (see Fig. 6a). The distance between SAA and the carboxylic acid group changes from 3.7 to 37.7 Å with an average of 16.5 Å. The interaction of SAA with the carboxylic acid is non-negligible in the first quarter of the sampling trajectory only, followed by a very weak interaction. Focusing on this period, SAA binds to the carboxylic acid with distance varies from 3.7 to 10.5 Å with an average of 5.6 Å. Moreover, the average binding energy of SAA with the carboxylic acid is -8.6 kJ/mol that is very close to what was obtained for SAA-naphthalene complex. Here, SAA is surrounded by 9.7 water molecules while the carboxylic acid group is surrounded by 10.0 water molecules.

Figure 6c shows that in model **Ib** SAA is strongly bound to the naphthalene functional group as well as the SOM cavity along the whole trajectory compared to interaction of SAA with free naphthalene (**Ia**, see Fig. 6a). Along the trajectory both SAA and naphthalene move out of the SOM cavity plane without dissociation for SAA (see Fig. S2a in the supplementary material). This indicates the substantial effect of the SOM cavity for binding of hydrophilic organic compounds. Due to this effect, the distance between SAA and naphthalene fluctuates from 3.8 to 15.3 Å with an average of 6.9 Å, which is relatively small compared to that in case of **Ia**. Along the whole trajectory, there is a relatively strong interaction for SAA with naphthalene and the SOM cavity. For the first half period of the trajectory, interaction of SAA with the SOM cavity is two times stronger than that with naphthalene. For the rest of the trajectory, there is a competition between the binding energy values for SAA-naphthalene and SAA-cavity interactions. This means that sometimes SAA-naphthalene interaction overcomes SAA-cavity interaction followed by overcome of SAA-cavity interaction and so on. The average binding energy for SAA with the whole model (**Ib**) is -37.0 kJ/mol. The binding energy of SAA with naphthalene is -12.9 kJ/mol to which the van der Waals interaction contributes with -10.8 kJ/mol while the contribution of the Coulomb interaction is -2.1 kJ/mol. SAA binds stronger to the SOM cavity with -24.1 kJ/mol where the contribution of the van der Waals interaction is -20.4 kJ/mol. In presence of the SOM cavity, solvation of water molecules of the interacting species is decreased due to shielding of the cavity. Hence, it is found that SAA is surrounded by 6.6 water molecules while naphthalene is surrounded by 5.1 water molecules. In comparison with **IIb**, SAA is highly mobilized during its interaction with the carboxylic acid functional group as well as the SOM cavity along the overall trajectory (Fig. 6d). SAA escapes the SOM cavity and appears mostly out of plane of the SOM cavity but the carboxylic acid fluctuates in the plane of the SOM cavity (see Fig. S2b in the supplementary material). Hence, the distance between SAA and carboxylic acid changes from 3.6 to 20.2 Å with an average of 10.3 Å. These are longer distances as compared to the SAA with **Ib** case. A weak interaction is found between SAA and the carboxylic acid functional group along some parts of this trajectory. On the other hand, a relatively moderate interaction is observed between SAA and the SOM cavity along the whole trajectory. The average binding energy for SAA with **IIb**, the carboxylic acid, and the SOM cavity are -21.1, -2.9, and -18.2 kJ/mol, respectively. This indicates that the cavity decreases the binding of SAA to the carboxylic acid compared to the free carboxylic acid functional group (**IIa**). Hence, 8.2 water molecules surround SAA while 11.4 water molecules surround the carboxylic acid. This means that changing of the naphthalene with the carboxylic acid group increases solvation of the complex sub-units that decreases the interaction strength. Thus, including the cavity will shield the carboxylic acid group from water but increase solvation of SAA.

Regarding model **Ic**, SAA is bound to the naphthalene functional group and part of the SOM cavity around the naphthalene functional group (Fig. 6e). During the trajectory, both SAA and naphthalene move out of plane of the SOM cavity (see Fig. S2c in the supplementary material). This leads to dissociation of the SAA-naphthalene complex followed by migration of SAA from the cavity to the water vicinity. Here, the distance between SAA and naphthalene alters from 3.7 to 21.1 Å with an average of 9.4 Å. Along the first two-thirds of the trajectory, there is relatively strong interaction for SAA with naphthalene compared to that with the SOM cavity. Vanishing of SAA-naphthalene interaction has been explored along the rest of the trajectory. A relatively strong interaction between SAA and a segment of the SOM cavity is found along the whole trajectory. Analysis of the first two-thirds of the trajectory yields distance between SAA and naphthalene fluctuates around 5.5 Å with maximum distance of 10.1 Å. The average binding energy for SAA with this whole model (**Ic**) is -21.3 kJ/mol in which -15.5 and -5.8 kJ/mol correspond to SAA-naphthalene and SAA-cavity interactions, respectively. This indicates that increasing of the cavity size decreases binding of SAA to the SOM cavity but increases binding of SAA to naphthalene. The binding of SAA to naphthalene in the model **Ic** is increased compared to the free naphthalene model **Ia**. Hence, SAA is surrounded by 8.3 water molecules while naphthalene is surrounded by 6.0 water molecules. SAA and naphthalene in case of the large cavity (**Ic**) are less solvated than that in case of absence of the cavity (**Ia**) but are more solvated than that in case of the small cavity (**Ib**). This leads to an intermediate binding energy that lies between energies obtained for the models with a small cavity and without any cavity. For SAA with model **IIc**, Fig. 6f and Fig. S2d in the supplementary material show that SAA interacts very weakly with the carboxylic acid functional group as well as the large SOM cavity. Along the trajectory, dissociation of SAA-naphthalene complex occurs followed by the removal of SAA from the cavity. Along the whole trajectory, the distance between SAA and the carboxylic acid fluctuates around 29.5 Å with minimum and maximum of 3.9 Å and 49.0 Å, respectively. Very weak interactions for SAA with the carboxylic acid and the SOM cavity are explored along the trajectory. As a result, dissociation of SAA complexes with the carboxylic acid and the SOM cavity take place at some points of the trajectory. Focusing on the periods of significant interaction yields an average distance between SAA and the carboxylic acid of 7.9 Å. Moreover, the average binding energy for SAA with **IIc** is -12.4 kJ/mol. The binding energy for SAA to the carboxylic acid is -2.4 kJ/mol while that to the SOM cavity is -10.0 kJ/mol. This indicates a larger cavity size diminishes the binding of SAA to the SOM cavity without significant effect on the carboxylic acid. The binding of SAA to the carboxylic acid is decreased compared to the free carboxylic acid model **IIa** but the cavity outweighs this effect yielding a higher binding energy than the free one. SAA is surrounded by 9.4 water molecules while the carboxylic acid is surrounded by 12.2 water molecules. Thus, SAA is less solvated than in case of no cavity but is more solvated than in case of the small cavity. Further, it was found that the carboxylic acid is more solvated than that in case of absence of the cavity as well as in case of the small cavity. Alike to the case of a large cavity including naphthalene, solvation procedure will lead to intermediate binding energy between in case of small cavity and absence of the cavity.

**Figure 7**

## 4 Summarizing Discussion

In summary, we have introduced a new SOM model and demonstrated how it can be used to describe the interaction of organic pollutants with SOM. The focus has been on two organic pollutants (HCB and SAA) which differ in their polarity. Experimental evidence was presented which showed that the interaction of SAA with SOM depends more on the chemical composition of SOM than on the SOM content. Moreover, it was confirmed that SAA obeys a site-specific adsorption on the soil surfaces. Furthermore, sorption hysteresis was observed indicating an inclusion of SAA inside soil voids. In addition, it was found that HCB is adsorbed stronger than SAA on similar soil sample. This indicated a preference of adsorption for hydrophobic organic compounds as compared to hydrophilic ones.

Molecular dynamics simulations have been performed to gain insight into the details of the interaction of these organic pollutants with SOM. Here, modeling of SOM has been introduced in terms of free SOM functional groups as well as SOM cavities (mimicking voids) containing the same functional groups. Two different, hydrophilic and hydrophobic, functional groups (carboxylic acid and naphthalene) as well as two cavities with different sizes were considered. Comparing the free SOM functional group and the one included into a cavity one can differentiate in simple terms adsorption and absorption processes for organic pollutants.

As a general result it can be stated that the hydrophobicity of the organic pollutant is a key property when it comes to binding to SOM, i.e. organic pollutants with high hydrophobicity bind stronger to SOM. This has been shown for HCB (highly hydrophobic organic compound) with SOM cavities and SOM functional groups where a binding was observed which is stronger than for SAA (lowly hydrophobic organic compound). For example, the average binding energy for SAA with naphthalene **Ia** equals to only about one third of that for HCB with naphthalene **Ia**. In addition, SAA is less bound than HCB to the carboxylic acid **IIa** which has half of the average binding energy value for HCB with **IIa**. Further, the average binding energy for SAA with the model **Ib** is around one half of that for HCB with **Ib.** In general, due to hydrophilicity of SAA, replacing HCB with SAA produces highly solvated complexes. This indicates that the binding tendency of HCB to SOM, via adsorption and/or absorption, is more pronounced than of SAA, which is in agreement with experiments (binding energies differ by factor 1.7-3.3). In fact the SOM-SAA is easily dissociated in contrast to the SOM-HCB one. For interaction of both HCB and SAA with SOM, the hydrophobic interaction (van der Waals interaction) exceeds the electrostatic interaction. Due to polarity of SAA, contribution of the latter interaction for binding of SAA to SOM is higher than that for HCB.

The presence of the SOM-cavity has a great effect on binding of the hydrophilic as well as the hydrophobic pollutants to SOM. For both SOM functional groups, HCB and SAA bind to the SOM model with small cavities (**Ib** and **IIb**) stronger than to the SOM model with large cavities (**Ic** and **IIc**) stronger than to the free systems (**Ia** and **IIa**). This indicates that the binding energy increases as the cavity size fits to the size of the pollutant. Therefore, increasing the number of SOM voids will increase binding of pollutants to SOM. The presence of the small and the large cavities has increased binding of HCB to SOM 2.5 and 1.7-2.2 times, respectively as compared to absence of the cavities. For SAA the respective increase is by 2.8-4.6 and 1.5-2.8 times. Thus, it can be concluded that although HCB binds to all SOM models stronger than SAA, the latter is more affected by the presence of the cavity. Moreover, one can translate the models of small and large cavities to models of low and high OC content, respectively. So in this case, it can be concluded that both HCB and SAA bind to the model of low OC content stronger than to the model of high OC content. This means that binding of HCB and SAA to SOM depends on the quality of SOM more than the quantity which comes in a good agreement with the experimental outcomes.

Further, it was found that HCB and SAA bind to the hydrophobic functional group (naphthalene) stronger than to the hydrophilic one (carboxylic acid) for most of SOM models. For models without a cavity, SAA binds to carboxylic acid stronger than to naphthalene. This indicates that the binding process (adsorption and/or absorption) occurs through a site-specific interaction that agrees with the experimental results. For details, HCB binds to naphthalene 1.7 times stronger compared to the carboxylic acid functional group in case of models with and without small cavity. In case of large cavity, naphthalene increases the binding energy of HCB 3.4 times compared to the carboxylic acid functional group. For the SOM models without cavity, binding of SAA to naphthalene is a little bit smaller than its binding to the carboxylic acid functional group. The presence of a cavity will increase binding of SAA to naphthalene 1.7 times as compared to the carboxylic acid functional group. Both HCB and SAA bind to the SOM models in the order of small SOM cavity containing naphthalene (**Ib**) > large SOM cavity containing naphthalene (**Ic**) > small SOM cavity containing carboxylic acid group (**IIb**) > large SOM cavity containing carboxylic acid group (**IIc**) > free naphthalene (**Ia**) ≥ free carboxylic acid group (**IIa**). This means that the effect of the SOM cavity exceeds that of the SOM functional group. This indicates to the significant role for the cavity, that is mimicking soil voids, in the binding process. This can support the experimental consequences regarding inclusion of the organic pollutants inside soil voids.

## 5  Acknowledgment

This Project was funded by the Deanship of Scientific Research (DSR) King Abdulaziz University, Jeddah, under Grant No. RG/18/34. In addition, this contribution is considered as a continuation for work that was supported by the Interdisciplinary Faculty (INF), University of Rostock, Germany. Therefore, The authors would like to acknowledge with thanks DSR and INF support for Scientific Research.18

# Tables

Table 1: The general composition of the different soil samples as well as the Freundlich and the linear adsorption parameters.

| Soil sample | | | | | SAA | | | | | | HCB[a] | | |
|---|---|---|---|---|---|---|---|---|---|---|---|---|---|
| | pH | C | N | C/N | Adsorption | | | Desorption | | | Adsorption | | |
| | $CaCl_2$ | % | | | $K_F$ | N | $r^2$ | $K_F$ | n | $r^2$ | $K_F$ | n | $r^2$ |
| U | 6.3 | 1.01 | 0.08 | 12.63 | 0.97 | 0.81 | 0.85 | 5.44 | 0.53 | | 4.02 | 0.75 | 0.99 |
| StmII | 6.4 | 1.23 | 0.05 | 26.71 | 2.17 | 0.69 | 0.95 | 6.44 | 0.44 | | | | |
| StmI | 6.6 | 1.45 | 0.11 | 13.18 | 1.15 | 0.82 | 0.92 | 2.41 | 0.91 | | | | |

[a] Data from Ahmed et al. (2014a).

Table 2: The calculated binding energies (in kJ/mol) of HCB with the sub-units of the different SOM models in Fig. 5. Both contributions from Lennard Jones (LJ) and Coulomb (COUL) interactions are given. The red values correspond to the average binding energies along the trajectory neglecting the dissociation periods.

| SOM model | SOM cavity | | SOM functional group | | Total energy of the model |
|---|---|---|---|---|---|
| | LJ | COUL | LJ | COUL | |
| **Ia** | -------- | -------- | -25.44 | -0.70 | -26.14 |
| **IIa** | -------- | -------- | -10.50 | 0.05 | -10.55 |
| | | | **-14.96** | **0.08** | **-14.88** |
| **Ib** | -36.73 | -0.05 | -27.68 | -0.45 | -64.91 |
| **IIb** | -35.07 | -0.13 | -3.23 | -0.03 | -38.46 |
| **Ic** | -24.32 | -0.18 | -19.56 | -0.29 | -44.34 |
| **IIc** | -30.20 | -0.16 | -2.66 | 0.01 | -33.03 |
| | | | **-3.13** | **0.01** | **-33.48** |

Table 3: The distances (in Å) between HCB and naphthalene or carboxylic acid functional group in the different SOM models as well as number of water molecules solvating HCB in a sphere of 5 Å.

| SOM model | Distance between HCB and SOM functional group [Å] | | | | no. of water around HCB |
|---|---|---|---|---|---|
| | Minimum | Maximum | Average | Standard deviation | |
| **Ia** | 3.29 | 9.34 | 4.61 | 1.22 | 4.5 |
| **IIa** | 3.54 | 47.15 | 12.42 | 11.05 | 5.7 |
| **Ib** | 3.39 | 10.59 | 4.54 | 1.18 | 1.6 |
| **IIb** | 3.46 | 13.74 | 7.78 | 2.02 | 3.8 |
| **Ic** | 3.39 | 10.48 | 5.81 | 1.83 | 3.7 |
| **IIc** | 3.57 | 19.95 | 8.85 | 3.66 | 4.1 |

Table 4: The calculated binding energies (in kJ/mol) of SAA with the sub-units of the different SOM models in Fig. 6. Both contributions from Lennard Jones (LJ) and Coulomb (COUL) interactions are given. The red values correspond to the average binding energies along the trajectory neglecting the dissociation periods.

| SOM model | SOM Cavity | | SOM functional group | | Total energy of the model |
|---|---|---|---|---|---|
| | LJ | COUL | LJ | COUL | |
| **Ia** | -------- | -------- | -3.26 **-8.40** | -0.03 **-0.08** | -3.29 **-8.48** |
| **IIa** | -------- | -------- | -3.26 **-7.01** | -0.75 **-1.57** | -4.0115 **-8.58** |
| **Ib** | -20.43 | -3.66 | -10.76 | -2.11 | -36.96 |
| **IIb** | -15.29 | -2.86 | -1.49 **-2.25** | -0.65 **-0.60** | -20.29 **-21.00** |
| **Ic** | -5.26 **-5.28** | -0.48 | -10.66 **-14.82** | -0.71 | -17.11 **-21.29** |
| **IIc** | -5.15 **-8.98** | -0.58 **-1.00** | -0.23 **-2.21** | -0.05 **-0.17** | -6.01 **-12.36** |

Table 5: The distances (in Å) between SAA and naphthalene or carboxylic acid functional group in the different SOM models as well as number of water molecules solvating HCB in a sphere of 5 Å.

| SOM model | Distance between SAA and SOM functional group [Å] | | | | no. of water around SAA |
|---|---|---|---|---|---|
| | Minimum | Maximum | Average | Standard deviation | |
| **Ia** | 3.61 | 40.39 | 20.61 | 11.75 | 9.6 |
| **IIa** | 3.68 | 37.66 | 16.50 | 9.06 | 9.7 |
| **Ib** | 3.78 | 15.28 | 6.87 | 2.70 | 6.6 |
| **IIb** | 3.62 | 20.19 | 10.31 | 4.10 | 8.2 |
| **Ic** | 3.68 | 21.14 | 9.39 | 5.54 | 8.2 |
| **IIc** | 3.92 | 48.99 | 29.55 | 11.13 | 9.4 |

# Figures

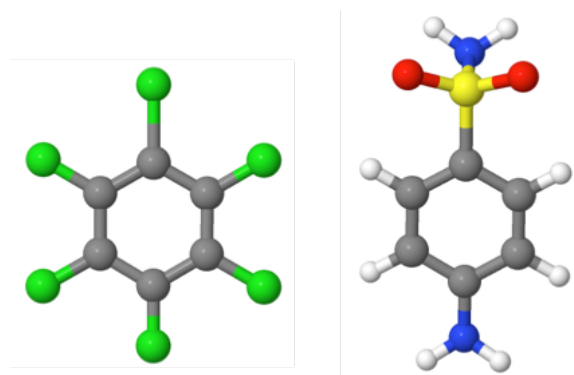

Figure 1: The pollutants considered in this study: hexcholorobenze (HCB, left) and sulfanilimide (SAA, right).

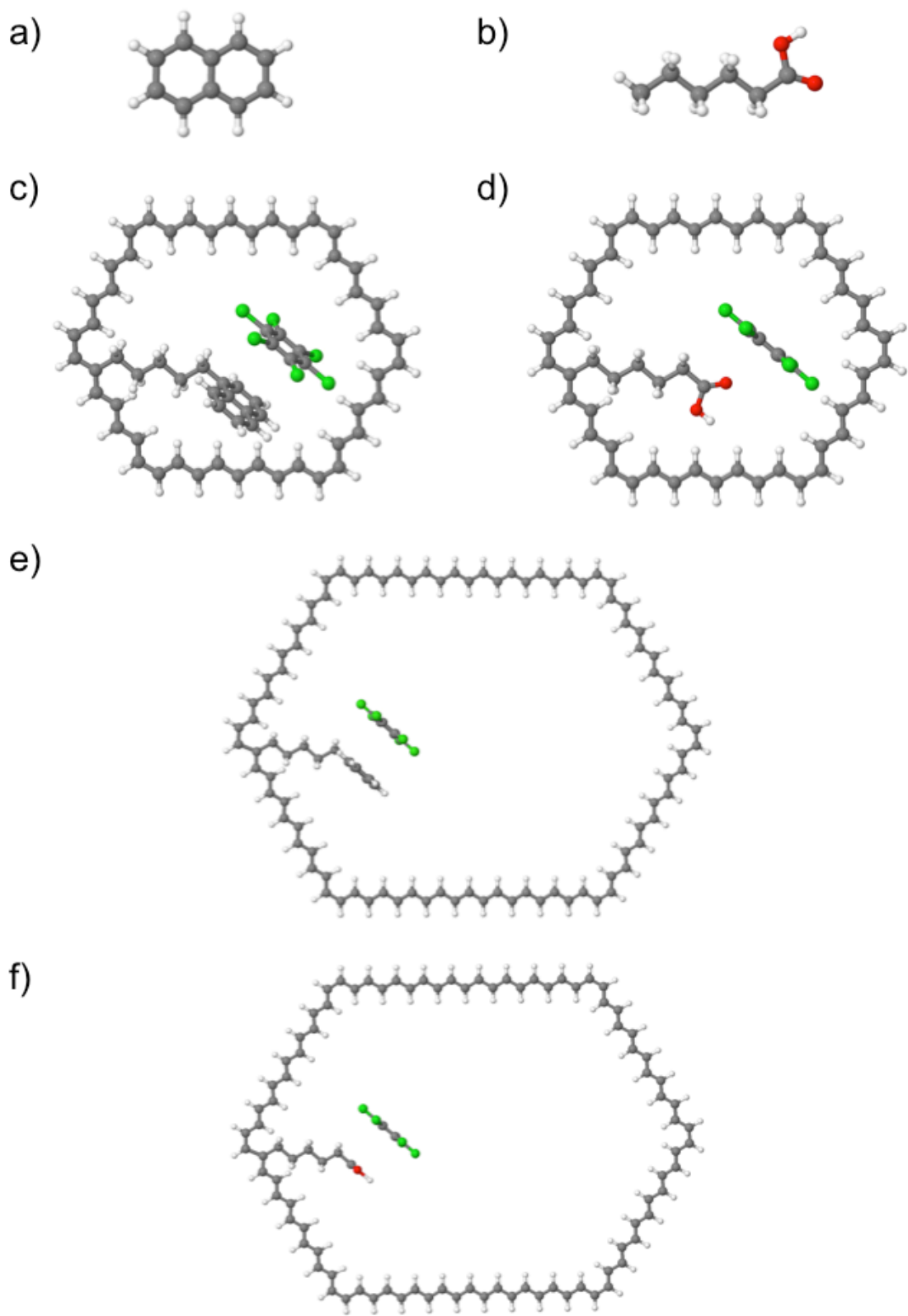

Figure 2: The considered SOM models: a) free naphthalene (**Ia**), b) free carboxylic acid group (**IIa**), c) small SOM cavity containing naphthalene (**Ib**), d) small SOM cavity containing carboxylic acid group (**IIb**), e) large SOM cavity containing naphthalene (**Ic**), and f) large SOM cavity containing carboxylic acid group (**IIc**). In addition it is shown how the pollutant, for example HCB, is placed into the cavity.

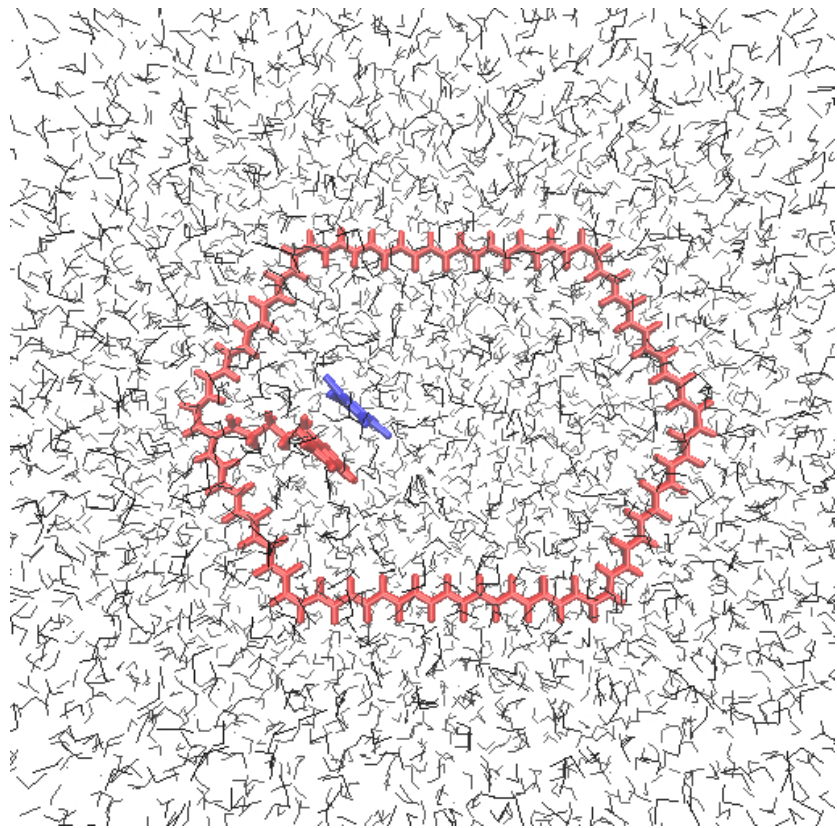

Figure 3: Snapshot of the equilibrated solvated complex of HCB (blue) with **Ic** (red) in water (black).

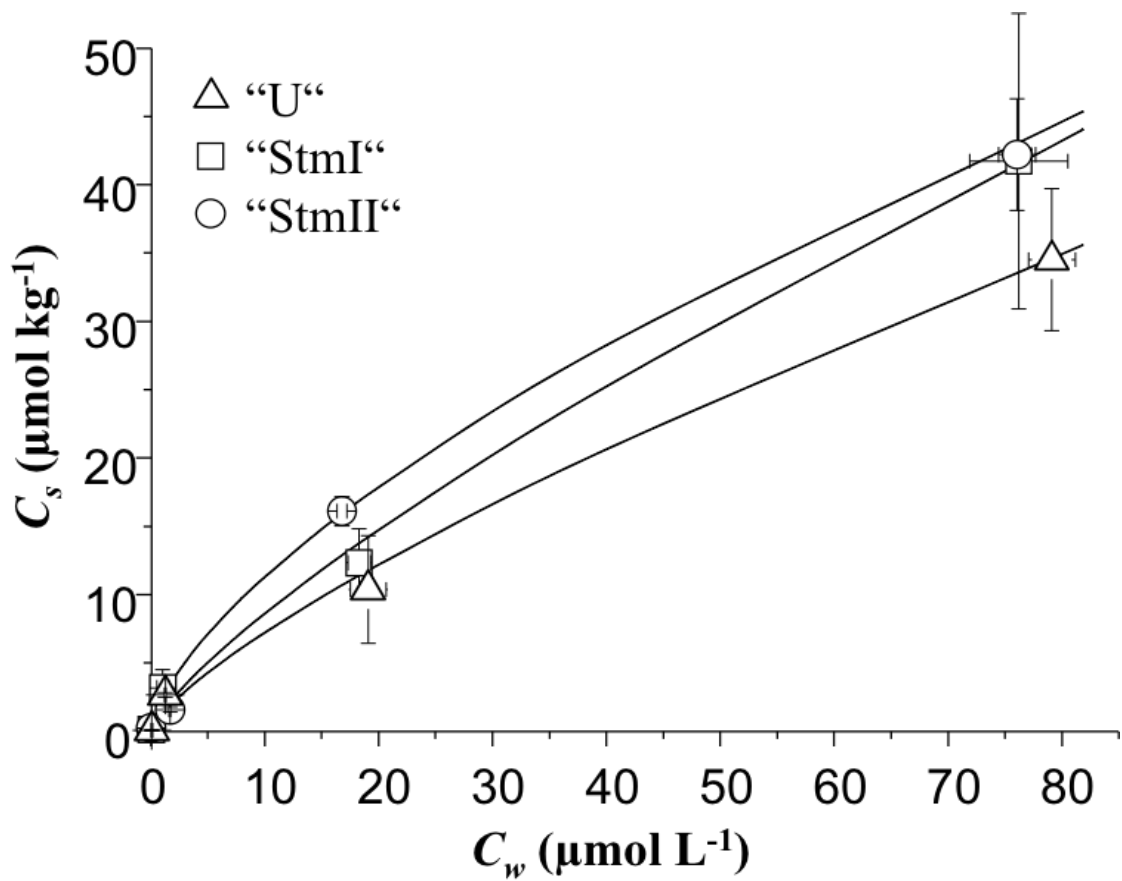

Figure 4: Adsorption of sulfanilamide (SAA) to different topsoils of the long-term fertilization experiment „Eternal Rye" at Halle, Germany. Symbols represent experimental data with standard error (bars), lines represent curve-fits modelled with the Freundlich equation.

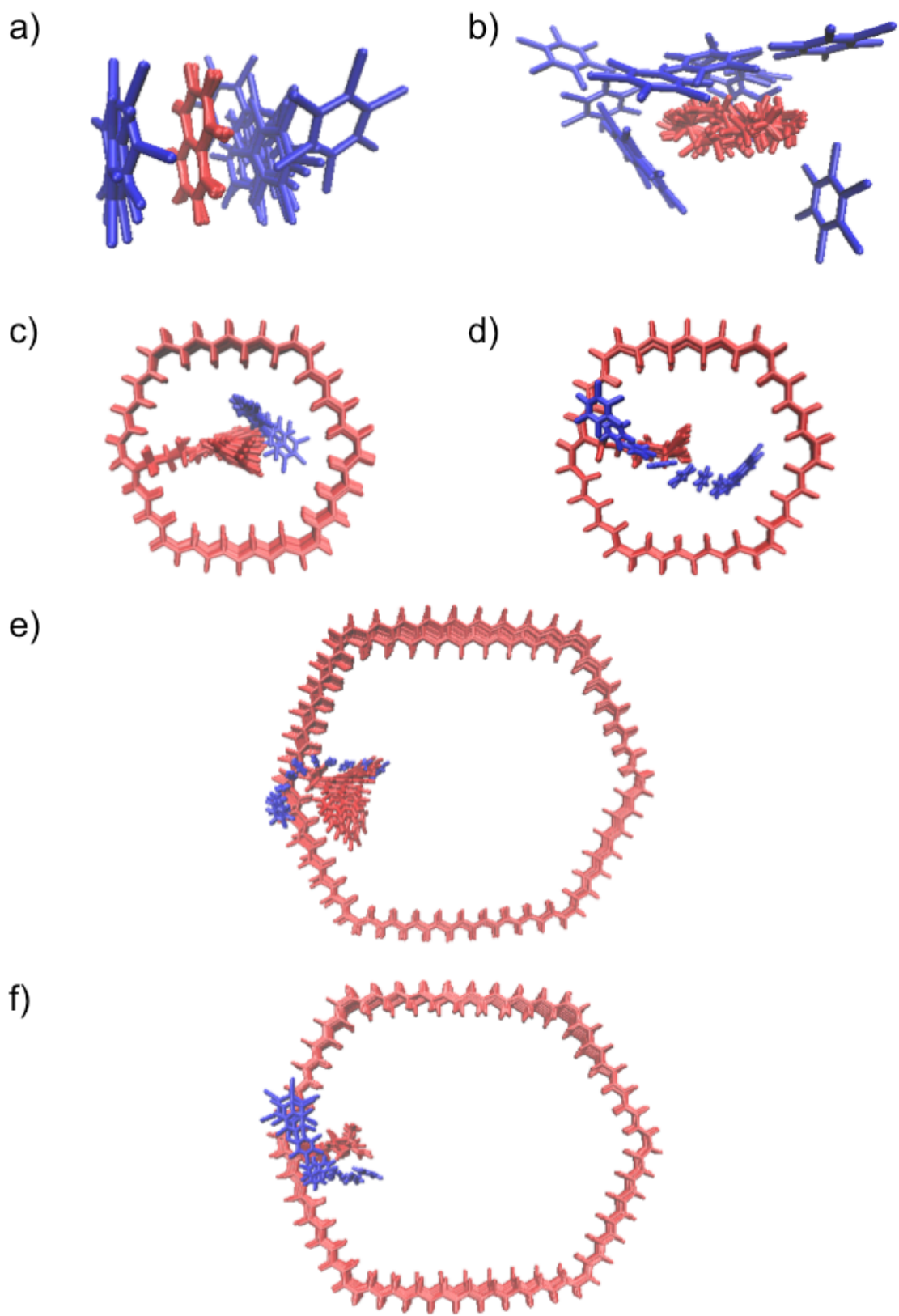

Figure 5: Overlay of eleven selected structures along an equilibrium trajectory for HCB (blue) with the different SOM models (red) in which a) corresponds to the free naphthalene (**Ia**), b) corresponds to the free carboxylic acid group (**IIa**), c) corresponds to the small SOM cavity containing naphthalene (**Ib**), d) corresponds to the small SOM cavity containing carboxylic acid group (**IIb**), e) corresponds to the large SOM cavity containing naphthalene (**Ic**), and f) corresponds to the large SOM cavity containing carboxylic acid group (**IIc**). Note that for visual clarity, an alignment of the pollutant-SOM structures has been performed by taking the SOM models as references.

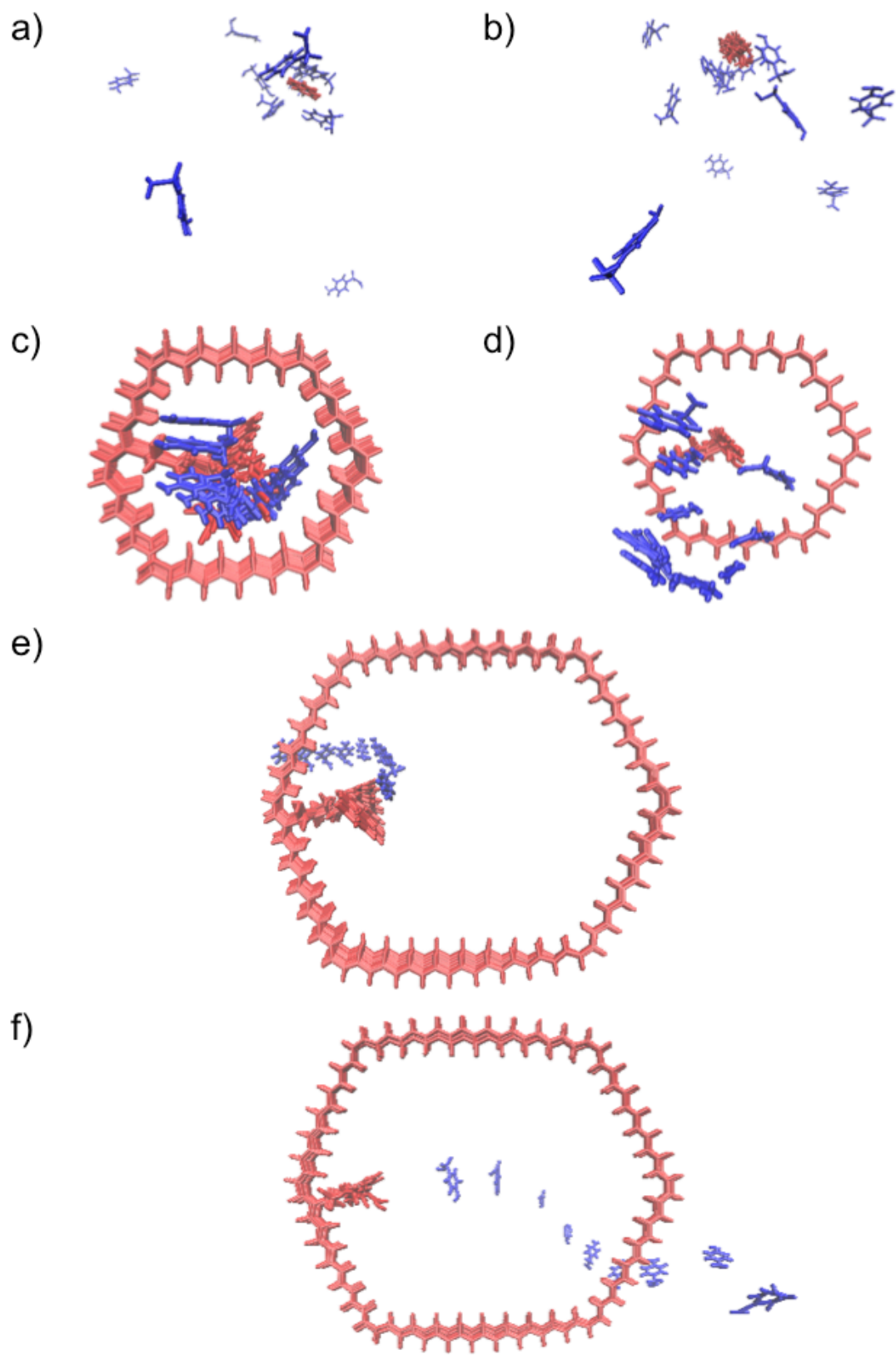

Figure 6: Eleven selected overlay structures along an equilibrium trajectory for SAA (blue) with the different SOM models (red) in which a) corresponds to the free naphthalene (**Ia**), b) corresponds to the free carboxylic acid group (**IIa**), c) corresponds to the small SOM cavity containing naphthalene (**Ib**), d) corresponds to the small SOM cavity containing carboxylic acid group (**IIb**), e) corresponds to the large SOM cavity containing naphthalene (**Ic**), and f) corresponds to the large SOM cavity containing carboxylic acid group (**IIc**). Note that for clarity of presentation, alignment of the pollutant-SOM structures has been performed by taking the SOM models as references.

# Supplementary Material

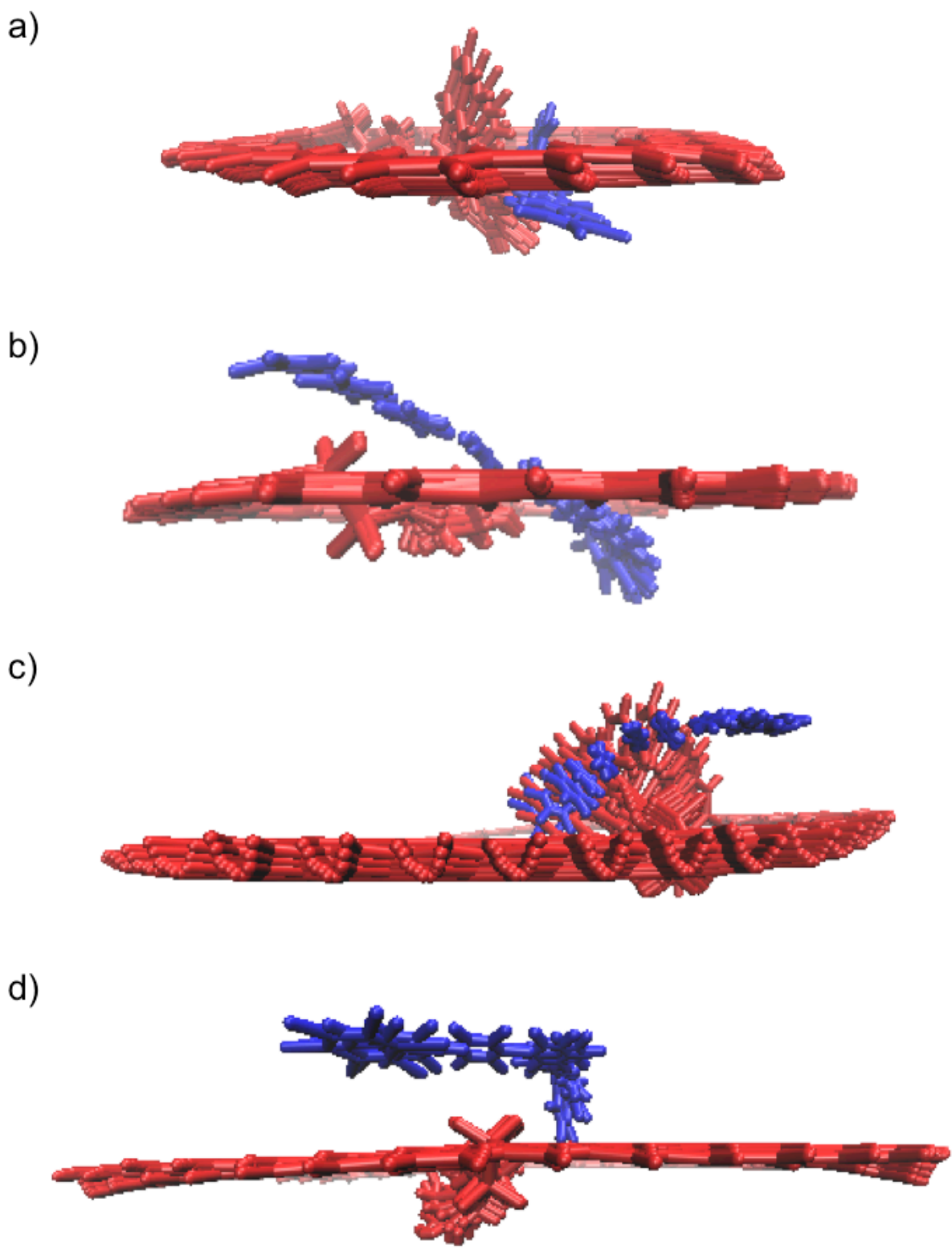

Figure S1: Overlay of eleven selected structures along an equilibrium trajectory for HCB (blue) with the modeled SOM, in side view of the cavity plane, (red) in which a) corresponds to the small SOM cavity containing naphthalene (Ib), b) corresponds to the small SOM cavity containing carboxylic acid group (IIb), c) corresponds to the large SOM cavity containing naphthalene (Ic), and d) corresponds to the large SOM cavity containing carboxylic acid group (IIc). Note that for visual clarity, an alignment of the pollutant-SOM structures has been performed by taking the SOM models as references.

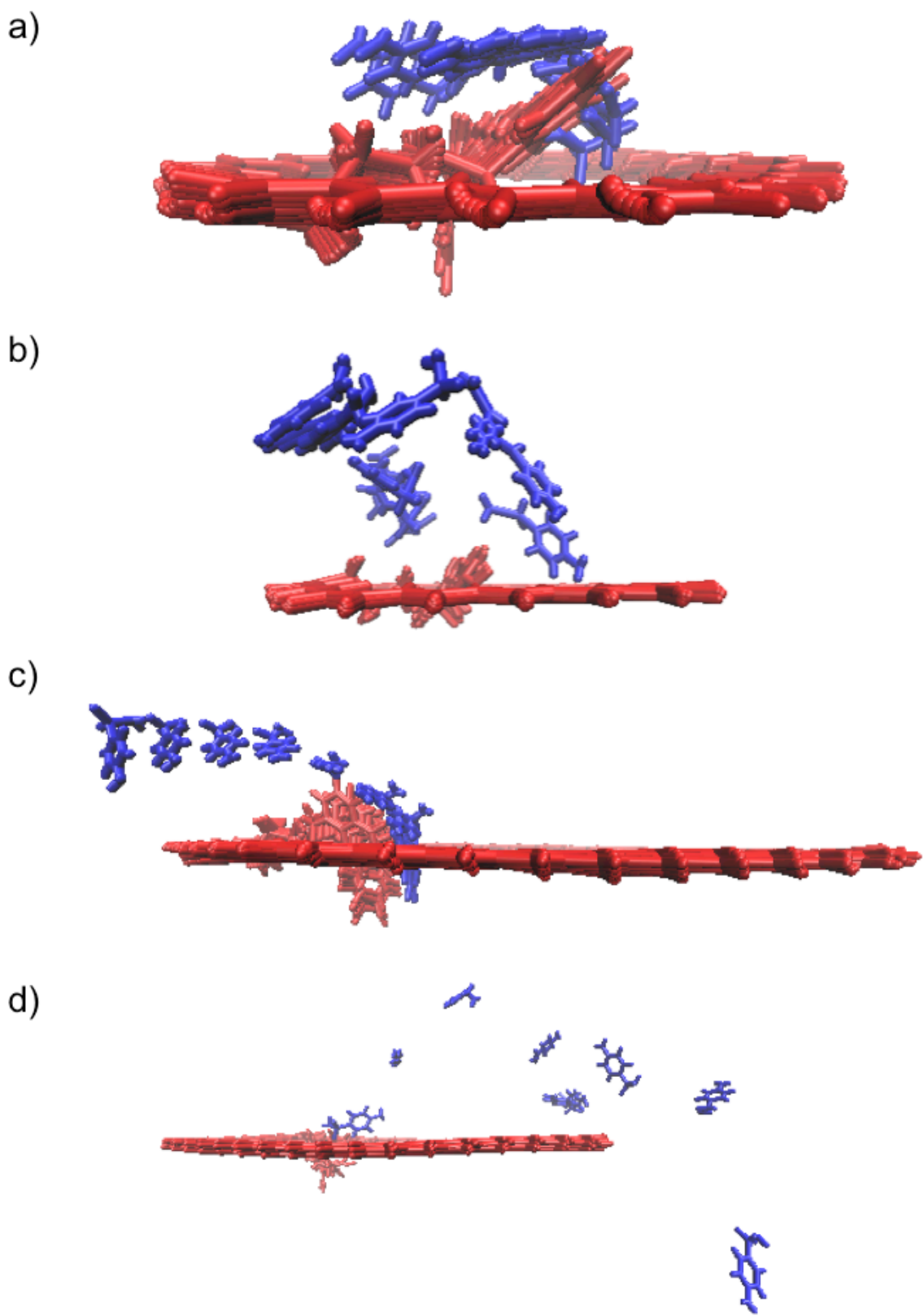

Figure S2: Overlay of eleven selected structures along an equilibrium trajectory for SAA (blue) with the moldeled SOM, in side view of the cavity plane, (red) in which a) corresponds to the small SOM cavity containing naphthalene (Ib), b) corresponds to the small SOM cavity containing carboxylic acid group (IIb), c) corresponds to the large SOM cavity containing naphthalene (Ic), and d) corresponds to the large SOM cavity containing carboxylic acid group (IIc). Note that for visual clarity, an alignment of the pollutant-SOM structures has been performed by taking the SOM models as references.